\documentclass[aps,prx, reprint,amsfonts,amsmath,amssymb,longbibliography,groupedaddress]{revtex4-2}

\usepackage{braket,bm,graphicx}

\begin{document}


\title{Mirror-symmetry-protected dynamical quantum phase transitions in topological crystalline insulators}



	\author{Ryo Okugawa}
\affiliation{%
	Graduate School of Information Sciences, Tohoku University, Sendai 980-8579, Japan
}%

\author{Hiroki Oshiyama}
\affiliation{%
	Graduate School of Information Sciences, Tohoku University, Sendai 980-8579, Japan
}%

\author{Masayuki Ohzeki}
\affiliation{%
	Graduate School of Information Sciences, Tohoku University, Sendai 980-8579, Japan
}%
\affiliation{%
	Institute of Innovative Research, Tokyo Institute of Technology, Yokohama 226-8503, Japan
}%
\affiliation{%
	Sigma-i Co., Ltd., Tokyo 108-0075, Japan
}%

\date{\today}

\begin{abstract}
    Dynamical quantum phase transitions (DQPTs) are topologically characterized in quantum quench dynamics in topological systems.
	In this paper, we study Loschmidt amplitudes and DQPTs in quantum quenches in mirror-symmetric topological phases.
	Based on the topological classification of mirror-symmetric insulators,
	we show that mirror symmetry creates symmetry-protected DQPTs.
	If mirror symmetry is present, topologically robust DQPTs can occur in quantum quenches, even in high-dimensional time-reversal invariant systems.
	Then, we also show that symmetry-protected DQPTs occur in quenches in two-dimensional chiral-symmetric systems with mirror symmetry.
	Mirror-symmetry-protected DQPTs can be easily captured by a reduced rate function.
	Moreover, we introduce dynamical topological order parameters for mirror-symmetry-protected DQPTs.
	Finally, we demonstrate DQPTs
	using lattice models for a time-reversal invariant topological crystalline insulator and a higher-order topological insulator.
\end{abstract}


\maketitle

	\section{Introduction}
	Topology plays an important role in the characterization of quantum phases
	in condensed matter physics \cite{Hasan10, Qi11, Goldman16, Chiu16}.
	Using topology, quantum phases can be systematically classified based on symmetry.
	Recent works have intensively studied topological phases protected by crystal symmetry,
	which are called topological crystalline phases \cite{Chiu16, Ando15}.
	For instance, mirror symmetry creates various novel phases
	\cite{Teo08, Hsieh12, Zhang13, Ueno13, Chiu13, Morimoto13, Shiozaki14, Okugawa18}.
	This concept has been extended to higher-order topological phases, which host topologically protected corner and hinge states 
	\cite{Benalcazar17S, Benalcazar17B, Fukui18, Langbehn17, Geier18, Khalaf18, Schindler18, Liu19, Imhof18, Serra-Garcia18, Mittal19}.
	
	Meanwhile, the idea of topological characterization has been applied to quench dynamics
	\cite{Caio15, Hu16, Wilson16, Caio16, Wang17, Sun18, Tarnowski19, Yang18, Zhang18, Unal20, Gong18, McGinley18, McGinley19, Nag19, Yang20, Mizoguchi21}. 
	Within a quantum quench, while a system is prepared as the ground state of an initial Hamiltonian,
	the state evolves under a different Hamiltonian with suddenly switched parameters. 
	A dynamical quantum phase transition (DQPT) is also known as a nonequilibrium topological phenomenon induced by quantum quenches
	\cite{Heyl18R, Zvyagin16, Vajna15, Schmitt15, Huang16, Budich16, Sharma16, Bhattacharya17, Heyl17, Bhattacharya17B, Sedlmayr18, Wang18, Bhattacharjee18, Qiu18, Zhou18, Mendl19, Jafari19, Lahiri19, Maslowski20, Hu20, Mishra20, Sadrzadeh21, Jafari21}.

	While a systematic understanding of phase transitions far from equilibrium is challenging,
	the concept of DQPTs can introduce a nonequilibrium counterpart
	of equilibrium phase transitions in quantum many-body dynamics \cite{Zvyagin16, Heyl18R}.
	In quenched systems, DQPTs emerge through zeros of the Loschmidt amplitude 
	\begin{align}
		&G(t)=\bra{\psi _0}{e^{-iH^ft}}\ket{\psi _0},
	\end{align}
	where $\ket{\psi _0}$ is the ground state of the initial Hamiltonian, and $H^f$ is the final Hamiltonian to evolve the state.
	DQPTs occur when the evolved state is orthogonal to the initial state.
	Because DQPTs are described as zeros of the Loschmidt amplitude,
	they can be considered analogous to the Fisher zeros of canonical partition functions by extending real time to complex time \cite{Heyl13, Heyl14}.
	Thus, if Fisher zeros cross the real time axis in the thermodynamic limit, DQPTs can occur.
	Similarly, as a nonequilibrium analog of free energy,
	one can introduce the (full) rate function $f(t)=-\lim _{N\rightarrow \infty}\log |G(t)|^2/N$ with system size $N$.
	Therefore, the rate function shows nonanalytic behavior at critical times.
	Moreover, dynamical topological order parameters (DTOPs) can be defined to characterize the real time dynamics
	\cite{Budich16, Sharma16, Heyl17, Bhattacharya17}.
	DTOPs change the value at critical times, making them  useful as order parameters for DQPTs. 
	Furthermore, DQPTs have been experimentally realized in many platforms
	\cite{Jurcevic17, Flaschner18, Guo19, Wang19, Tian19, Yang19, Xu20}.
	
    DQPTs are related to equilibrium topological phase transitions.
	Previous works revealed that DQPTs in band insulators are predictable
	if quenches cross topological phase transitions in two-dimensional (2D) class A and one-dimensional (1D) class AIII \cite{Vajna15, Huang16}.
	However, the conditions of DQPTs are still elusive.
	Since class A (AIII) includes only magnetic (sublattice-symmetric) insulators,
	the previous theory cannot be applied to typical time-reversal invariant insulators.
	Moreover, crystal symmetry enriches topological phases in lattice systems.
	The conditions of DQPTs should also vary by crystal symmetries.
	Therefore, we reveal the conditions in other symmetry classes and dimensions.

	In this paper, we propose DQPTs topologically characterized by mirror symmetry,
	and clarify the conditions in some mirror-symmetric classes.
	By mirror symmetry, we can topologically predict DQPTs even if time-reversal symmetry is present. 
	The robust DQPTs can appear in topological crystalline insulators with mirror symmetry.
	We also investigate DTOPs to characterize DQPTs in mirror-symmetric topological phases.
	
	The remainder of this paper is organized as follows.
	In Sec.~\ref{topoDQPT}, we define symmetry-protected DQPTs, 
	and reveal how they are realized by nontrivial band topology in mirror-symmetric classes.
	We also discuss DTOPs for DQPTs protected by mirror symmetry.
	In Sec.~\ref{latticemodel}, we demonstrate symmetry-protected DQPTs in lattice models with mirror symmetry.
	Finally, we present our conclusions in Sec.~\ref{CandD}.
	
	\section{DQPTs and mirror symmetry} \label{topoDQPT}
	We show that symmetry-protected DQPTs can be realized from nontrivial topology by mirror symmetry.
	Hereinafter, we use superscripts $i$ and $f$ to represent the initial and final states, respectively.
	
	\subsection{DQPTs in topological systems}
	First, we briefly review DQPTs in 2D class A and 1D class AIII 
	before detailing our results.
	
	\subsubsection{DQPTs and band topology}
	We introduce symmetry-protected DQPTs in two-band insulators according to Ref.~\onlinecite{Vajna15},
	although they can be extended to multiband systems with one occupied band \cite{Huang16, Mendl19}.
	We consider a two-band model parameterized by a vector $\bm{d}(\bm{k})=(d_x, d_y, d_z)$.
	The Hamiltonian is generally written as
	\begin{align}
		H=\sum _{\bm{k}}\bm{c}^{\dagger}_{\bm{k}}[\bm{d}(\bm{k})\cdot \bm{\sigma}] \bm{c}_{\bm{k}}, \label{twoband}
	\end{align}
	where $\bm{\sigma}=(\sigma _{x}, \sigma_{y},\sigma _{z})$ are Pauli matrices for the internal degrees of freedom A and B, and
	$\bm{c}^{\dagger}_{\bm{k}}=(c^{\dagger}_{\bm{k}A}, c^{\dagger}_{\bm{k}B})$ are fermionic creation operators.
	
	Because of the translation invariance,
	the ground state consists of eigenstates $\ket{\psi _{\bm{k}}}$ at each momentum $\bm{k}$.
	Then, the Loshcmidt amplitude can be written as $G(t)=\prod _{\bm{k}}G_{\bm{k}}(t)$ with
	\begin{align}
		G_{\bm{k}}(t)=\cos (|\bm{d}^f(\bm{k})|t)
		+i\hat{\bm{d}^i}(\bm{k}) \cdot \hat{\bm{d}^f}(\bm{k})\sin (|\bm{d}^f(\bm{k})|t), \label{twobandLA}
	\end{align}
	where $\hat{\bm{d}}^{i(f)}$ is the normalized vector $\bm{d}^{i(f)}$ for the initial (final) Hamiltonian.
	The product in $G(t)$ is taken for all wavevectors in the Brillouin zone (BZ).
	If we extend real time $t$ to complex time in Eq.~(\ref{twobandLA}),
	we can obtain the Fisher zeros on the complex plane as
	\begin{align}
		z_n(\bm{k})=\frac{i\pi}{|\bm{d}^f|}\left( n+\frac{1}{2}\right) -\frac{1}{|\bm{d}^f|}\mathrm{arctanh}[\hat{\bm{d}^i}\cdot \hat{\bm{d}^f}],
		~n \in \mathbb{Z}. \label{zeros}
	\end{align}
	Thus, if $\hat{\bm{d}^i}\cdot \hat{\bm{d}^f}=0$ is satisfied at critical momenta $\bm{k}=\bm{k}_c$, 
	the Fisher zeros appear at real times, which leads to DQPTs through nonanalyticities of the rate function.

	We can predict DQPTs from the topological difference between pre-quench and post-quench two-band Hamiltonians 
	if the class is characterized by a Chern or winding number \cite{Vajna15}.
	As explained below, the DQPTs are robust
	because their occurrence is guaranteed only by symmetry and topology of the initial and final states.
	Therefore, such DQPTs are called topological or symmetry-protected DQPTs.

	In 2D class A, a Chern number $C$ topologically classifies insulators without any symmetries.
	In Appendix \ref{appti},
	we give an expression for the Chern number for the two-band model.
	Importantly, we can always obtain momenta with $\hat{\bm{d}^i}\cdot \hat{\bm{d}^f}=\pm 1$ 
	for quenches between two phases with different absolute values of the Chern numbers.
	Hence, because $\hat{\bm{d}^i}\cdot \hat{\bm{d}^f}=0$ is allowed on a curve in the BZ by the continuity,
	DQPTs necessarily occur in the quench dynamics if $|C^i| \neq |C^f|$.
	
	1D class AIII gapped systems can be also topologically characterized by a winding number $\nu$ (see also Appendix \ref{appti}).
	The Hamiltonian has chiral symmetry given by $SH(\bm{k})S^{-1}=-H(\bm{k})$ with a unitary matrix $S$ whose square is $+1$.
	Similar to class A, when the winding number of the initial state is different from that of the final state,
	$\hat{\bm{d}^i}\cdot \hat{\bm{d}^f}=0$ can be satisfied in the 1D BZ.
	Therefore, we can find DQPTs at critical times when $\nu ^i \neq \nu ^f$ in a quench.
	
	We also comment on the difference between DQPTs in the two classes.
	In 2D systems, the Fisher zeros can form areas on the complex plane
	since $z_n(\bm{k})$ in Eq.~(\ref{zeros}) depends on $k_x$ and $k_y$.
	Especially, if a Chern number for two-band insulators characterizes DQPTs,
	the Fisher zeros form areas leading to nonananlyticities in the first derivative of the full rate function \cite{Vajna15, Schmitt15}.
	Namely, nonananlyticities do not appear in the full rate function.
	Because the Chern number describes 2D band topology,
	we cannot see the nonanalytic behavior in 1D topological systems.
	
	\subsubsection{Dynamical vortices and DTOPs}
	Topologically protected DQPTs can be captured by dynamical vortices and DTOPs \cite{Budich16, Sharma16, Heyl17, Bhattacharya17}.
	We also introduce the Pancharatnam geometric phase to analyze the DQPTs.
	In the form of $G_{\bm{k}}(t)=|G_{\bm{k}}(t)|e^{i\phi _{\bm{k}}(t)}$, 
	the geometric phase $\phi _{\bm{k}}^G(t)$ is defined as
	\begin{align}
		\phi _{\bm{k}}^G(t) = \phi _{\bm{k}}(t) -\phi _{\bm{k}}^{D}(t),
	\end{align}
	where $\phi _{\bm{k}}^{D}(t)=-\int _0^t ds \bra{\psi _{\bm{k}}(s)}H^f\ket{\psi _{\bm{k}}(s)}$ is a dynamical phase.
	While the dynamical phase is continuous in time,
	the Pancharatnam geometric phase shows discontinuity at critical momenta
	when topological DQPTs occur.
	This discontinuity appears as a $\pi$-phase jump
	because $\phi _{\bm{k}_c}^{D}(t)$ vanishes, and $G_{\bm{k}_c}(t)$ changes the sign \cite{Budich16, Sharma16, Heyl17, Bhattacharya17}.
	Moreover, $\phi _{\bm{k}}^G(t)$ is fixed to zero at all times
	if the momentum satisfies $\hat{\bm{d}^i} \cdot \hat{\bm{d}^f}=+1$ or $-1$.
	Hence, a critical momentum lies on any path that connects two fixed points with $\hat{\bm{d}^i} \cdot \hat{\bm{d}^f}=+1$ and $-1$.
	
	In 2D two-band systems, a $\pi$ jump yields vortex-antivortex pairs in the profile of the geometric phase \cite{Heyl17, Qiu18}.
	If the first Fisher zero gives rise to a DQPT, vortices and antivortices are created in pairs.
	These pairs move on the curves of the critical momenta 
	while real time is in the area of Fisher zeros because of the 2D topological aspect.
	When a DQPT occurs again at the edge of the area of Fisher zeros,
	the vortices and antivortices are annihilated.
	Thus, dynamical vortices are regarded as signatures of DQPTs.
	
	Furthermore, we can use DTOPs to identify topologically protected DQPTs accompanied by a $\pi$ jump \cite{Budich16, Qiu18}.
	A DTOP is an integer winding number defined as
	\begin{align}
		\nu _{D}(t)=\frac{1}{2\pi}\int _{l_D}\nabla _{\bm{k}}\phi _{\bm{k}}^G(t)\cdot d\bm{k}.
	\end{align}
	Here, we can choose a contour connecting two fixed points as $l_D$. 
	In 1D two-band systems, the value of $\nu _{D}(t)$ changes only when DQPTs topologically occur.
	Therefore, DTOPs serve to observe DQPTs in quenches between topologically different phases.

	\subsection{Mirror-symmetry-protected DQPTs}
	Using mirror symmetry,
	we generalize the previous topological approach to explore DQPTs in other symmetry classes and dimensions,
	which we name mirror-symmetry-protected DQPTs.
	{For the realization,
	we consider topological crystalline insulators whose band topology are characterized by mirror symmetry.}
	We show that the symmetry-protected DQPTs occur in the quenched topological crystalline insulators.
	This paper focuses on four-band insulators with two occupied bands,
	which can include time-reversal invariant systems with Kramers degeneracy.
	
	First, we clarify the relationship between mirror symmetry and Loschmidt amplitude.
	Mirror symmetry is generally represented by a unitary matrix $M$ such that
	\begin{align}
		MH(\bm{k})M^{-1}=H(\bm{k}_{\parallel},-k_{\perp}), \label{mirrorsymm}
	\end{align}
	where $H(\bm{k})$ is a mirror-symmetric Hamiltonian in the BZ, and $\bm{k}_{\parallel} ~(k_{\perp})$ represents the components of the wave vector $\bm{k}$ invariant (flipped) under the mirror operation.
	Although $M^2=+1$ or $-1$,
	we focus on the case where $M$ is Hermitian for $M^2=1$ in this paper.
	Whereas $M^2=-1$ in spinful systems, we can redefine the mirror symmetry by adding a phase factor such that it becomes Hermitian.
	
	Mirror-symmetric Hamiltonians can be block-diagonalized into two mirror sectors $H_{\pm}(\bm{k})$
	with mirror eigenvalues $\pm 1$ on the mirror-invariant lines and planes.
	We consider quantum quenches that retain mirror symmetry.
	Therefore, we can factorize the Loschmidt amplitude further.
	For mirror-invariant momenta, we have
	\begin{align}
		G_{\bm{k}}(t)=G_{\bm{k},+}(t)G_{\bm{k},-}(t), \label{Gmirror}
	\end{align}
	where $G_{\bm{k}, \pm}(t)$ are Loschmidt amplitudes given by $H_{\pm}(\bm{k})$.
	Consequently, nonanalytic behaviors of the rate function arise from each mirror sector.
	
	Hereinafter, we investigate how mirror symmetry enriches DQPTs in quantum quenches.
	From the factorization in Eq.~(\ref{Gmirror}), the Fisher zeros can be understood from the band topology of the mirror sectors.
	In general, topological classification of mirror sectors $H_{\pm}(\bm{k})$ can be different from that of $H(\bm{k})$
	because the mirror sectors do not necessarily have the same symmetries as $H(\bm{k})$ \cite{Teo08, Hsieh12, Chiu13, Morimoto13, Shiozaki14}.
	In other words, even if $H(\bm{k})$ has nonspatial symmetries, including time-reversal symmetry,
	each of the mirror sectors may or may not keep them.
	This is determined by whether operators of nonspatial symmetries commute or anticommute with a mirror operator \cite{Chiu13, Morimoto13, Shiozaki14}.
	Accordingly, mirror sectors can acquire different topological invariants in some symmetry classes.
	Hence, we can realize DQPTs topologically
	if the Chern or winding number can be introduced for mirror sectors.
    For instance, such nontrivial band structures are realizable in topological crystalline phases.
	In this case, the critical momenta can be found on the mirror-invariant momenta.
	In this paper, when DQPTs can be characterized by a Chern number or winding number for mirror sectors in such a manner,
	we call them mirror-symmetry-protected DQPTs, and otherwise accidental DQPTs.
	In particular, we study DQPTs in class AII and class AIII 
	because the classes can realize topological crystalline insulators with the nonzero Chern or winding number.
	
	In class AII, we can indeed obtain nonzero Chern numbers $C_{\pm}$ for mirror sectors $H_{\pm}(\bm{k})$ 
	when the time-reversal operator anticommutes with the mirror operator.
	In Appendix \ref{appti}, we also explain the band topology for classes A and AIII with mirror symmetry.
	Thus, nonanalytic behavior from $G_{\bm{k},\pm}(t)$ is predictable from the absolute values of nonzero Chern numbers $C_{\pm}$.
	Because $C_++C_-=0$ by time-reversal symmetry,
	such systems are topologically classified by a topological invariant $C_M=(C_+-C_-)/2$,
	which is called a mirror Chern number \cite{Teo08, Hsieh12}.
	Hence, we can characterize DQPTs by the absolute value of a mirror Chern number, despite time-reversal symmetry.
	Unless mirror symmetry is broken,
	DQPTs are robust for quantum quenches between the trivial phase and the topological crystalline phase with a nonzero mirror Chern number.
	
	Moreover, other winding numbers can be introduced to 2D class AIII with mirror symmetry if the chiral operator commutes the mirror operator.
	It should be noted that the 2D strong topology is trivial in the absence of mirror symmetry \cite{Chiu13, Morimoto13}.
	As with class AII, we can define winding numbers on the mirror-invariant line;
	therefore, a mirror winding number $\nu _M=(\nu _+-\nu _-)/2$ specifies topological phases \cite{Zhang13, Imhof18}.
	Hence, when a system is quenched between two phases with different mirror winding numbers,
	mirror-symmetry-protected DQPTs can occur.
	
	In addition, we compare mirror-symmetry-protected DQPTs with other topologically robust DQPTs in high-dimensional systems.
	When a system can be seen as a collection of 1D (2D) topological systems,
	robust DQPTs also occur because winding numbers (Chern numbers) can be defined for general lines (planes) in the BZ \cite{Vajna15, Schmitt15, Lahiri19}.
	The band topology originates from translational symmetry.
	In contrast, the origin of mirror-symmetry-protected DQPTs is the band topology of the mirror-invariant line or plane.
	A mirror Chern number (winding number) is related to dynamical Fisher zeros only on the mirror-invariant plane (line).
	Although DQPTs can occur from critical momenta at generic points in a quench between a trivial phase and a topological crystalline phase,
	they are accidental in general.
	Therefore, mirror-symmetry-protected DQPTs are different from nonanalyticities from a collection of lower-dimensional topological systems.
	
	\subsection{Mirror DTOPs and reduced rate function}
	We define DTOPs to distinguish mirror-symmetry-protected DQPTs from accidental ones.
	Unlike the two-band case, $G_{\bm{k}}(t)$ does not necessarily show a $\pi$ jump of the Pancharatnam geometric phase
	when mirror-symmetry-protected DQPTs occur.
	For example, no $\pi$ jump can be seen when $G_{\bm{k}_c,+}(t)=G_{\bm{k}_c,-}(t)$, as shown in Eq.~(\ref{Gmirror}).
	Thus, we must introduce DTOPs to reflect the topological properties of mirror sectors.
	
	First, we consider dynamical vortices on 2D mirror-invariant planes for signatures of mirror-symmetry-protected DQPTs.
	Mirror symmetry-protected DQPTs stem from the Loschmidt amplitudes $G_{\bm{k},\pm}(t)=|G_{\bm{k},\pm}(t)|e^{i\phi _{\bm{k, \pm}}(t)}$.
	Namely, in four-band topological crystalline insulators,
	we can independently treat the mirror sectors described as two-band Hamiltonians.
	Therefore, we can separately introduce the Pancharatnam geometric phases for the mirror sectors $H_{\pm}(\bm{k})$, defined as
	\begin{align}
		\phi _{\bm{k},\pm}^G(t) =\phi _{\bm{k},\pm}(t) - \phi _{\bm{k},\pm}^{D}(t).
	\end{align}
	The geometric phase $\phi _{\bm{k},+(-)}^G(t)$ shows a $\pi$ jump
	when the mirror sector $H_{+(-)}(\bm{k})$ contributes to DQPTs.
	Hence, we can diagnose mirror-symmetry-protected DQPTs by tracing the dynamical vortices in each mirror sector.
	
	Second, we also introduce DTOPs to mirror sectors on mirror-invariant lines.
	For 2D mirror-symmetric insulators, we define DTOPs as 
	\begin{align}
		\nu _{D}^{\pm}(t)=\frac{1}{2\pi}\int _{l_m}\nabla _{\bm{k}}\phi _{\bm{k},\pm }^G(t)\cdot d\bm{k},
	\end{align}
	where $l_m$ is a path that connects two different fixed points on the mirror-invariant line.
	We call $\nu _{D}^{\pm}(t)$ mirror DTOPs.
	If symmetry-protected DQPTs occur from either mirror sector,
	the mirror DTOP changes the value.

	Furthermore, we employ a rate function to clearly detect mirror-symmetry-protected DQPTs.
	In our case, a reduced rate function can be used to highlight DQPTs due to low-dimensional topology in the BZ.
	In Ref. \cite{Qiu18}, the reduced rate function was introduced to grasp nonanalyticities of a Loschmidt amplitude from a fraction of the BZ, defined as
	\begin{align}
		f_r(t) = -\frac{1}{2\pi}\oint _{l} d\bm{k}\log |G_{\bm{k}}(t)|^2.
	\end{align}
	A reduced rate function can visualize DQPTs as nonanalyticities at critical times
	because it is a line integral in the BZ \cite{Qiu18}.
	The loop $l$ is essentially chosen to pass through two different fixed points.
	In the 2D plane, we can determine $l$ as a loop in which none of dynamical vortices are related by discrete symmetries \cite{Qiu18}.
	Because such a loop can pass through critical momenta for edges of the Fisher zero areas for DQPTs, 
	$f_r(t)$ can precisely determine critical times.
	Therefore, we utilize a reduced rate function on the mirror-invariant lines and planes to see mirror-symmetry-protected DQPTs.
	The loop can be chosen according to discrete symmetry of the mirror sectors. 

	Since a reduced rate function $f_r(t)$ is a line integral,
	it is more accessible than the full rate function described as an integration over the entire BZ \cite{comment2}.
	Mirror-symmetry-protected DQPTs occur in more than one dimension.
	Hence, $f_r(t)$ is helpful to analyze the critical times
	because one only investigates the time evolution of wavefunctions at specific momenta in the BZ experimentally.

	\section{DQPTs in lattice models} \label{latticemodel}
	In this section, we verify mirror-symmetry-protected DQPTs in 2D lattice systems.
	To do so, we study a topological crystalline insulator in class AII
	and a higher-order topological insulator in class AIII.
	The two models have four bands with mirror symmetry.
	We assume that the lowest two bands are occupied in both models to realize topological crystalline phases.
	
	\begin{figure*}[t]
		\includegraphics[width=17cm]{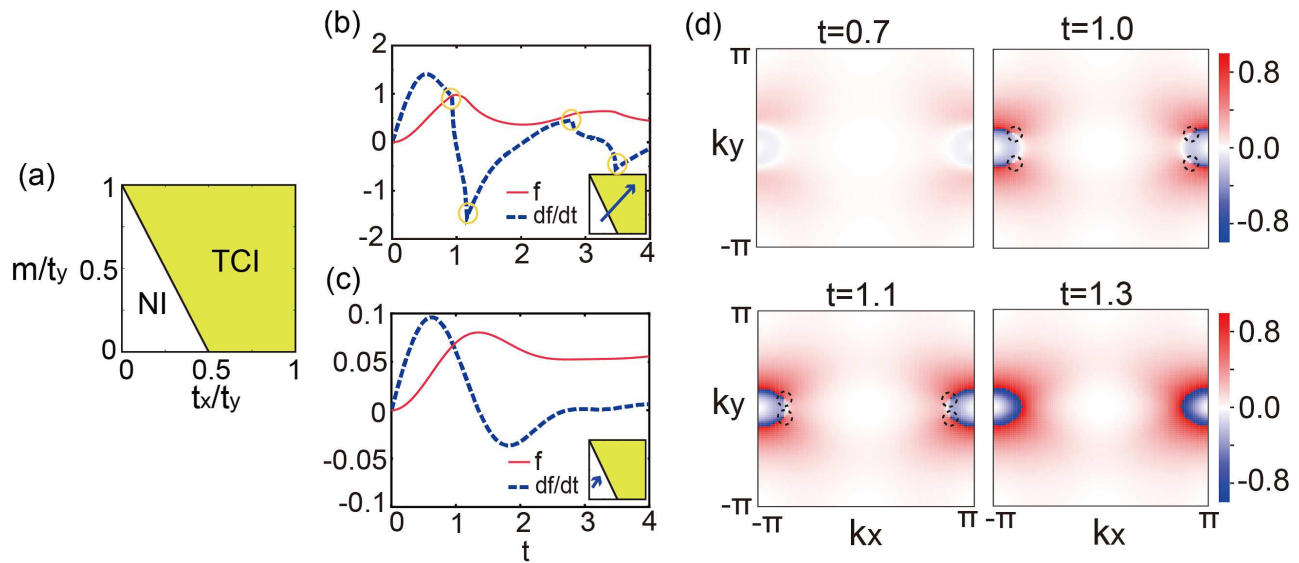}
		\caption{\label{TCIDQPTfull}(a) Phase diagram of the topological crystalline insulator model. 
			TCI represents a topological crystalline insulator phase with $C_M=-1$,
		    and NI represents a normal insulator phase that is topologically trivial.
		    (b) and (c) Full rate function $f(t)$ and first time derivative $df/dt$ for the topological crystalline insulator model.
		    The Hamiltonian is quenched 
		    from the NI phase with $(t_x/t_y,m/t_y)=(0.2,0.2)$ to the TCI phase with $(t_x/t_y,m/t_y)=(0.8,0.8)$ in (b) as well as
		    within the NI phase $(t_x/t_y,m/t_y)=(0.1,0.2)\rightarrow (0.2,0.4)$ in (c).
		    We fix the other parameters $v/t_y=v'/t_y=1.0$.	The unit of time is taken as $1/t_y$. 
			The circles in (b) indicate DQPTs found in the first time derivative of the full rate function. 
			(d) Phase profiles $\phi ^G_{\bm{k},+}(t)/\pi$ on the 2D BZ for (b). The dotted circles indicate dynamical vortices and antivortices.
		    The two pairs are connected by inversion symmetry of the mirror sector.
			}
	\end{figure*}	
	
	\subsection{Topological crystalline insulator with time-reversal symmetry}
	We consider a four-band model for a topological crystalline insulator with time-reversal symmetry to observe DQPTs protected by mirror symmetry.
	Mirror-symmetric topological crystalline insulators are characterized by a nonzero mirror Chern number.
    The topological crystalline insulators show gapless boundary states \cite{Teo08, Hsieh12}.
	To show DQPTs, we investigate the following Hamiltonian: 
	\begin{align}
		H^{\mathrm{TCI}}(\textbf{k})&=(m + t_x -t_x\cos k_x -t_y\cos k_y )s_0 \mu _z \notag \\ 
		&+v\sin k_x s_x\mu _x +v'\sin k_y s_0\mu _y, \label{3Dmodel}
	\end{align}
	where $s_{x,y,z} ~(\mu _{x,y,z})$ and $s_0 ~(\mu _0)$ are Pauli matrices and the identity matrix for spin (orbitals), respectively. 
	We set all the parameters to be positive. For simplicity, we assume that $m/t_y<1$ and $t_x/t_y<1$.
	This model has time-reversal symmetry $T=-i s_y\mu _0K$ with complex conjugation $K$.
	
	The band topology of the model can be characterized by mirror symmetry with respect to the $xy$ plane,
	which we write as $M_{xy}$.
	By choosing the mirror symmetries to be Hermitian,
	$M_{xy}$ is represented by
	\begin{align}
		&M_{xy}H^{\mathrm{TCI}}(\bm{k})M_{xy}^{-1}=H^{\mathrm{TCI}}(\bm{k}), ~M_{xy}=s_x\mu _0.
	\end{align}
	In this model, the 2D BZ is the mirror-invariant plane.
	Because the mirror symmetry satisfies $\{T, M_{xy}\} =0$,
	the system can have a nonzero mirror Chern number.

	To see the band topology, we block-diagonalize the model into mirror sectors.
	The mirror sectors with mirror eigenvalues $\pm 1$ are 
	\begin{align}
		H^{\mathrm{TCI}}_{\pm}(\bm{k}) 
		&= \pm v\sin k_x \mu _x+v'\sin k_y\mu _y \notag \\
		&+(m+t_x-t_x\cos k_x -t_y\cos k_y) \mu _z. \label{modelsector}
	\end{align}
	The two mirror sectors can be parametrized
	by two vectors, $\bm{d}_{\pm}^{\mathrm{TCI}}(\bm{k})=(\pm v\sin k_x, v'\sin k_y, m+t_x -t_x\cos k_x -t_y\cos k_y)$.
	The mirror sectors can be regarded as two inversion-symmetric Chern insulators with Chern numbers of opposite signs \cite{Hsieh12}.
	The phase diagram can be obtained from the band topology of the two mirror sectors [Fig.~\ref{TCIDQPTfull}(a)].
	In this model, the mirror sectors give $G_{\bm{k},+}(t)=G_{\bm{k},-}(t)$
	and have four fixed points $(0,0), (\pi ,0), (\pi, 0)$, and $(\pi, \pi)$.

	We investigate a quantum quench between the normal insulator phase and the topological crystalline insulator phase in the phase diagram.
	Figure ~\ref{TCIDQPTfull} (b) shows the rate function and the first time derivative
	for the quench that crosses the two different phases.
	In this case, nonanalyticities appear in the time derivative of the full rate function instead of the full rate function itself
	since dynamical Fisher zeros form areas on the complex plane for quenches
	in a Chern insulator \cite{Schmitt15, Vajna15}.
	Because of the nonzero mirror Chern number, we can see mirror-symmetry-protected DQPTs
	as nonanalyticities of the first time derivative of the full rate function.
	However, if the system is quenched within the normal insulator phase,
    DQPTs do not necessarily occur [Fig.~\ref{TCIDQPTfull} (c)].
	
	Moreover, we confirm dynamical vortices on the 2D BZ.
	Because $G_{\bm{k},+}(t)=G_{\bm{k},-}(t)$, 
	we compute the Pancharatnam geometric phase $\phi _{\bm{k},+}^G(t)$ for the mirror sector $H_+^{\mathrm{TCI}}(\bm{k})$.
	Figure \ref{TCIDQPTfull} (d) shows pair creation and annihilation of the dynamical vortices.
	The two vortex-antivortex pairs are created at the first DQPT.
	Then, the vortices move on the critical momenta symmetrically with respect to the origin.
	When the next DQPT occurs, all vortices and antivortices are annihilated.
	Therefore, we can see that the dynamical vortices reflect the mirror-symmetry-protected DQPTs.
	
\begin{figure}[t]
	\includegraphics[width=8.5cm]{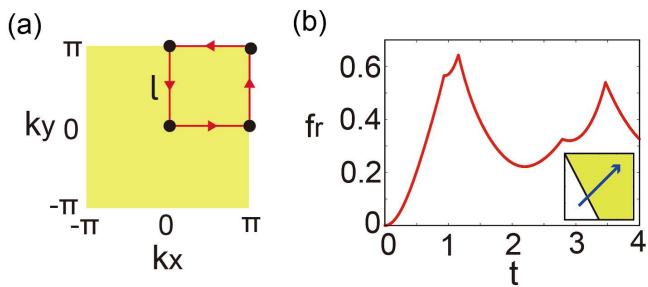}
	\caption{\label{TCIDQPT} (a) Loop $l$ for the reduced rate function on the BZ. The black points are the fixed points.
		(b) Reduced rate function for the topological crystalline insulator model. 
		The parameters are the same as those for the full rate function in Fig.~\ref{TCIDQPTfull} (b).
	}
	\end{figure}

	Here, we calculate a reduced rate function for the same quench.
	We emphasize that reduced rate functions manifest nonanalyticities at critical times
	because they can be effectively regarded as 1D rate functions.
	We can choose the path $l$ for the reduced rate function shown in Fig.~\ref{TCIDQPT} (a)
	because the mirror sectors are inversion-symmetric.
	We can find DQPTs in the reduced rate function in Fig.~\ref{TCIDQPT} (b)
	when the quench crosses the topological phase transition.	
	The critical times are consistent with those obtained for the full rate function.
	Therefore, we can see that mirror-symmetry-protected DQPTs can also be well understood
	with the reduced rate function on the appropriate loop.

	\subsection{Higher-order topological insulator with chiral symmetry}
	We also study quantum quenches in the Benalcazal-Bernevig-Hughes model with mirror symmetries \cite{Benalcazar17S, Benalcazar17B}.
	When the 2D model has chiral symmetry,
	zero-energy modes emerge at the corners as a manifestation of the higher-order topology \cite{Benalcazar17S, Benalcazar17B, Imhof18, Fukui18, Okugawa19B, Hayashi19}.
	The chiral-symmetric Hamiltonian has nonzero mirror winding numbers in the higher-order topological phase.
	Thus, topological nontriviality yields DQPTs for quenches between the trivial and the higher-order topological phases.
	
	The model Hamiltonian is given by
	\begin{align}
		H^{\mathrm{BBH}}(\bm{k})=(\gamma +\lambda \cos k_x)\sigma _x\tau _0 - \lambda \sin k_x\sigma _y\tau _z \notag, \\
		-(\gamma +\lambda \cos k_y)\sigma _y\tau _y-\lambda \sin k_y\sigma _y\tau _x, 
	\end{align}
	where $\lambda$ and $\gamma$ are real hopping parameters.
	Here, $\sigma _{x,y,z}$ and $\tau _{x,y,z}$ are Pauli matrices for the four sublattices, and $\sigma _0$ and $\tau _0$ are the identity matrices.
	The chiral symmetry is represented by $S=\sigma _z\tau _0 $.
	This model has two diagonal mirror symmetries which commute with the chiral symmetry.
	The two symmetries are described as $M_1H^{\mathrm{BBH}}(k_x,k_y)M_1^{-1}=H^{\mathrm{BBH}}(k_y, k_x)$
	and $M_2H^{\mathrm{BBH}}(k_x,k_y)M_2^{-1}=H^{\mathrm{BBH}}(-k_y, -k_x)$.
	The matrix representations are
	\begin{align}
		M_1&= \frac{1}{2}(\sigma _0 +\sigma _z)\tau _z + \frac{1}{2}(\sigma _0 -\sigma _z)\tau _x, \\
		M_2&= \frac{1}{2}(\sigma _0 +\sigma _z)\tau _x - \frac{1}{2}(\sigma _0 -\sigma _z)\tau _z, 
	\end{align}
	which satisfy $[M_1, S]=[M_2, S]=0$.
	Therefore, the mirror sectors on the mirror-invariant lines $k_x=\pm k_y$ have chiral symmetry.

\begin{figure}[t]
		\includegraphics[width=8cm]{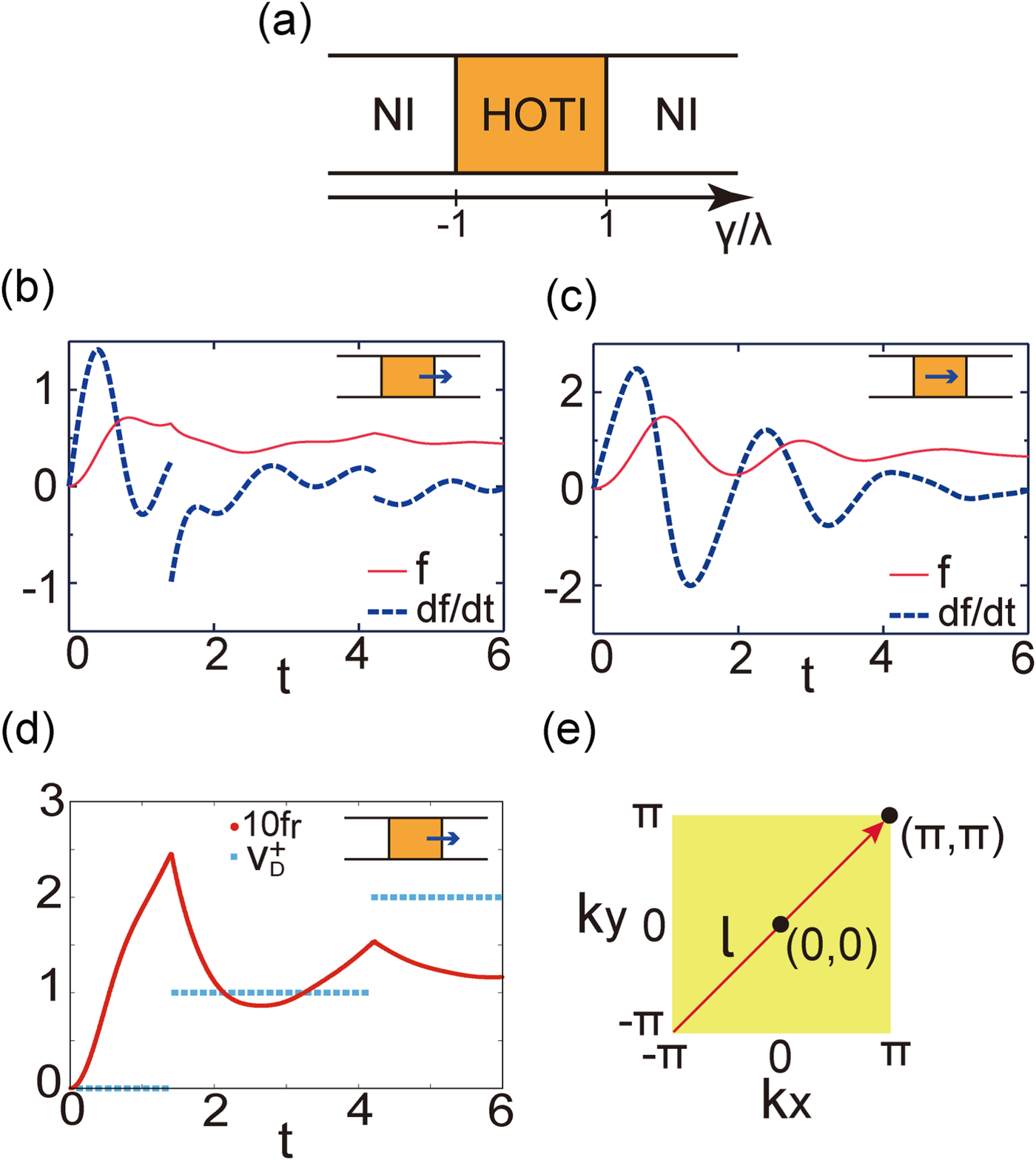}
		\caption{\label{HOTIDQPT} (a) Phase diagram for the Benalcazal-Bernevig-Hughes model.
			In (a), HOTI (NI) represents the higher-order topological insulator (normal insulator) phase with $\nu _M=-1 (0)$.
			(b) and (c) Full rate function $f(t)$ and first time derivative $df/dt$ for the model.
			The Hamiltonian is quenched from the HOTI phase ($\gamma/\lambda=0.5$) to the NI phase ($\gamma/\lambda=1.5$) in (c) as well as 
			within the HOTI phase ($\gamma/\lambda=-0.5\rightarrow 0.5$) in (d). The unit of time is taken as $1/\lambda$.
		    (d) Reduced rate function and mirror DTOP. 
		    The reduced rate function $f_r(t)$ is computed along the  mirror-invariant line $k_x=k_y$.
		    (e) Contour $l$ for the reduced rate function on the line $k_x=k_y$.
		    The two black points are the fixed points on the line.
		}
	\end{figure}

	We calculate winding numbers on the mirror-invariant line $k_x=k_y$ to consider DQPTs.
	This model also has fourfold rotational symmetry $C_4=[(\sigma _x+i\sigma _y)\tau _0-(\sigma _x-i\sigma _y)(i\tau _y)]/2$.
	Thus, we can focus on the mirror-invariant line $k_x=k_y$
	because the other line $k_x=-k_y$ contributes to DQPTs in the same manner.
	The mirror sectors with eigenvalue $\pm 1$ are written as
	\begin{align}
		H^{\mathrm{BBH}}_{\pm}(k)=\sqrt{2}
		\begin{pmatrix}
			0 & \gamma +\lambda e^{\pm ik} \\
			\gamma +\lambda e^{\mp ik} & 0
		\end{pmatrix}. \label{BBHsector}
	\end{align}
	The Hamiltonians can be parameterized by two vectors $\bm{d}_{\pm}^{\mathrm{BBH}}(k)
	=\sqrt{2}(\gamma +\lambda \cos k, \mp \lambda \sin k ,0)$.
	The winding numbers $\nu _{\pm}$ for $H_{\pm}^{\mathrm{BBH}}(k)$ become $\mp 1$
	when $|\gamma /\lambda| < 1$, which corresponds to the higher-order topological phase.
	The phase diagram is illustrated in Fig.~\ref{HOTIDQPT}(a).
	The higher-order topological phase has the nonzero mirror winding number $\nu _M=-1$.

	From the topological argument, we can find mirror-symmetry-protected DQPTs in quantum quenches in the Benalcazal-Bernevig-Hughes model.
	If the final and initial states have different band topology, DQPTs inevitably occur because of the symmetry protection.
	We compute the full rate function and the first time derivative for the model.
	When the system is quenched from the higher-order topological phase to the trivial phase,
	DQPTs can be seen in the full rate function [Fig.~\ref{HOTIDQPT} (b)].
	Meanwhile, DQPTs do not occur in the quench within the topological phase, as shown in Fig.~\ref{HOTIDQPT} (c).

	Here, we consider a topological origin of the mirror-symmetry-protected DQPTs by using a reduced rate function.
	Because zeros of the Loschmidt amplitudes are guaranteed by the band topology of the mirror-invariant lines,
	we compute the reduced rate function along the line $k_x=k_y$ [Fig.~\ref{HOTIDQPT} (e)].
	The calculation for the DQPTs is shown in Fig.~\ref{HOTIDQPT} (d).
	The critical times agree with those of the full rate functions. 
	Thus, we can see that the nonanalyticities are related to the band topology due to the mirror symmetry.
	Therefore, the reduced rate function is also beneficial for grasping mirror-symmetry-protected DQPTs.
	
	Moreover, we study mirror DTOPs on the $k_x=k_y$ line.
	The fixed points on the line $k_x=k_y$ are $(0,0)$ and $(\pi ,\pi)$.
	We calculate the mirror DTOP $\nu _+^D(t)$ for the mirror sector $H^{\mathrm{BBH}}_{+}(k)$
	because $G_{\bm{k},+}(t)=G_{\bm{k},-}(t)$ on the mirror-invariant line in this model.
	We choose the path for the DTOP to connect the two fixed points.
	Figure \ref{HOTIDQPT} (d) shows that $\nu _+^D(t)$ changes the value when mirror-symmetry-protected DQPTs occur.
	Hence, we can see that the mirror DTOP can characterize symmetry-protected DQPTs.
		
	\section{Conclusion and Discussion} \label{CandD}
	In this paper, we studied DQPTs protected by mirror symmetry
	and elucidated the relationship between the Loshcmidt amplitudes and the crystalline topology.
	We have shown 
	that mirror-symmetry-protected DQPTs can occur
	in quenches between a trivial phase and topological crystalline phase characterized by mirror symmetry.
	The nonanalytic behaviors of the Loshcmidt amplitudes can be predicted from the mirror Chern numbers and mirror winding numbers.
	Moreover, we proposed an approach using a reduced rate function to easily visualize mirror-symmetry-protected DQPTs.
	Mirror-symmetry-protected DQPTs can be analyzed with DTOPs and dynamical vortices in mirror sectors.
	Our work has revealed nonequilibrium phenomena unique to crystalline topology, including higher-order topology.

	Mirror-symmetry-protected DQPTs can be found in other mirror-symmetric topological phases. 
	For example, topological crystalline insulators are realizable
	not only in two-dimensional but also in three-dimensional systems \cite{Teo08, Hsieh12, Chiu13, Morimoto13}.
	Some three-dimensional second-order topological insulators are also characterized by a mirror Chern number
	\cite{Langbehn17, Geier18, Schindler18}.
	Additionally, some Weyl and Dirac semimetals can also have mirror Chern numbers on their mirror-invariant planes
	because the topological gapless phases can be regarded as intermediate phases between topologically different gapped phases
	\cite{Murakami07, Murakami08, Burkov11, Yang14, Okugawa14, Murakami17}.
	Therefore, if such topological systems are quenched, symmetry-protected DQPTs can occur.
	Furthermore, superconducting systems in some classes can be topologically classified by a Chern or winding number.
	The topological invariants also characterize DQPTs \cite{Vajna15}.
	Thus, the idea of mirror-symmetry-protected DQPTs can be extended to topological superconductors with mirror symmetry.
	Hence, mirror-symmetry-protected DQPTs can occur in quench dynamics in various topological systems.
	
	{We theoretically clarified topological aspects of DQPTs protected by mirror symmetry.
	Meanwhile, the possibility of experimental measurement of mirror DTOPs is important.
	Also, studying observables related to symmetry-protected DQPTs is helpful for the experimental detection.
	We leave them to future work.}
	
	\begin{acknowledgments}
		We thank M. Okuyama for their valuable discussions.
		This work was supported by JSPS Grant-in-Aid for Scientific Research on Innovative Areas ``Discrete Geometric Analysis for Materials Design" Grant No. JP17H06469.
	\end{acknowledgments}

	\appendix
	\section{Chern and winding numbers} \label{appti}
	We explicitly write topological invariants for two-band topological systems in Eq.~(\ref{twoband})
	to describe DQPTs in a self-contained way.
	In 2D class A, the Chern number for the occupied band can be represented by 
	\begin{align}
		C=\frac{1}{4\pi}\int _{\mathrm{BZ}}dk_xdk_y \hat{\bm{d}}\cdot (\partial _{k_x}\hat{\bm{d}}\times \partial _{k_y}\hat{\bm{d}}). \label{Chern}
	\end{align}
	We also give an expression of the winding number in 1D class AIII.
	When we take a basis where chiral symmetry is diagonal, we obtain $d_z=0$.
	Then, the winding number for the band below zero energy can be written as
	\begin{align}
		\nu = \frac{1}{2\pi}\int _{\mathrm{BZ}}dk(\hat{d}_x\partial _k\hat{d}_y-\hat{d}_y\partial _k\hat{d}_x). \label{winding}
	\end{align}
	{We can extend these topological invariants to topological crystalline insulators with mirror symmetry,
	as discussed below.
	In addition, if a bulk insulator has the above nonzero topological invariant,
	topologically protected edge states appear.
	Actually, various topological boundary states have been experimentally observed
	in crystalline solids \cite{Hasan10, Qi11, Ando15} and cold-atom systems \cite{Goldman16}.}
	
	{In this paper, we have studied topological crystalline insulators
	whose mirror sectors have nonzero topological invariants in Eqs. (\ref{Chern}) and (\ref{winding}).}
	We here explain how mirror-symmetric insulators in class AII can have nonzero mirror Chern numbers
	which characterize topological crystalline insulators.
	Because the class has time-reversal symmetry $T$ for $TH(\bm{k})T^{-1}=H(-\bm{k})$, any Chern number of the ground state will be zero \cite{Qi08}.
	Nevertheless, mirror symmetry $M$ can give nonzero Chern numbers
	to the mirror sectors in class AII \cite{Teo08, Hsieh12, Chiu13, Morimoto13, Shiozaki14}.
	The band topology is understandable as follows.
	Let us write the relation between $T$ and $M$ as $TM=\eta _TMT$ with $\eta _T=\pm 1$.
	When $\eta _T=1 (-1)$, $M$ and $T$ commute (anticommute).
	In the ground state, the Bloch state $\ket{u_{\bm{k}}}$ has time-reversal partner $T\ket{u_{\bm{k}}}$.
	On the 2D mirror-invariant planes,
	we can label the Bloch states as $\ket{u_{\bm{k}}, q}$ by the mirror eigenvalues $q= \pm 1$.
	The time-reversal partner $T\ket{u_{\bm{k}}, q}$ has mirror eigenvalue $\eta _Tq$
	because $M[T\ket{u_{\bm{k}}, q}]=\eta _T TM\ket{u_{\bm{k}}, q}=\eta _Tq T\ket{u_{\bm{k}}, q}$.
	If $T\ket{u_{\bm{k}}, q}$ has mirror eigenvalue $-q$,
	it is a Bloch state in a different mirror sector.
	Therefore, when $\eta _T=-1$, i.e., $TM=-MT$, no mirror sector has time-reversal symmetry.
	Thus, the mirror sectors can obtain nonzero Chern numbers.

	We also discuss 2D class AIII with mirror symmetry.
	Let us consider Bloch states on the mirror-invariant lines.
	Because the system has chiral symmetry $S$, we express the relation between $S$ and $M$ as $SM=\eta _SSM$ with $\eta _S=\pm 1$.
	When a Bloch state $\ket{u_{\bm{k}}, q}$ has energy eigenvalue $E$,
	$S\ket{u_{\bm{k}}, q}$ has energy $-E$ by chiral symmetry.
	Meanwhile, $S\ket{u_{\bm{k}}, q}$ has mirror eigenvalue $\eta _Sq$
	because $MS\ket{u_{\bm{k}}, q}=\eta _SSM\ket{u_{\bm{k}}, q}=\eta _SqS\ket{u_{\bm{k}}, q}$.
	To retain chiral symmetry in each mirror sector, the system must have $\eta _S=1$ for $q=\eta _Sq$.
	Therefore, if $SM=MS$ by $\eta _S=1$, each mirror sector has chiral symmetry on the 1D mirror-invariant line.
	Hence, we can introduce winding numbers for the mirror sectors.

\bibliography{DQPT}

\begin{thebibliography}{82}
\bibitem [{com()}]{comment2}%
  \BibitemOpen
  \href@noop {} {\bibinfo  {journal} {In $d$-dimensional band insulators, the
  full rate function is rewritten as $f(t)=-\frac{1}{(2\pi )^{d}}\int
  _{\mathrm{BZ}}d\bm{k}\log |G_{\bm{k}}(t)|^2$}\ }\BibitemShut {NoStop}%
\end{thebibliography}

\end{document}